\documentclass[twocolumn]{aastex7}

\accepted{\today}

\submitjournal{ApJ Letters}

\usepackage{amsmath}
\usepackage{comment}
\graphicspath{{./}{figures/}}

\shorttitle{Optical–Neutrino Test of SN 2023\lowercase{uqf}}
\shortauthors{Sawada, Inoue, \& Ashida}

\begin{document}

%%%%%%%%% TITLE PAGE --  AUTHORS -- %%%%%

\title{Interaction-powered Type Ibn Supernovae as a Transient PeVatron Candidate: The Case of SN 2023uqf}

\author[orcid=0000-0003-4876-5996,gname=Ryo, sname='Sawada']{Ryo Sawada} 
\affiliation{Center for Interdisciplinary Theoretical and Mathematical Sciences (iTHEMS), RIKEN, Saitama 351-0198, Japan}
\affiliation{Institute for Cosmic Ray Research, The University of Tokyo, Kashiwa, Chiba 277-8582, Japan}
\email[show]{ryo.sawada@riken.jp}

\author[orcid=0009-0004-3148-0462,gname=Yusuke, sname='Inoue']{Yusuke Inoue} 
\affiliation{Department of Astronomy, Kyoto University, Kitashirakawa-Oiwake-cho, Sakyo-ku, Kyoto, 606-8502, Japan}
\email[]{yusuke@kusastro.kyoto-u.ac.jp}

\author[orcid=0000-0003-4136-2086,gname=Yosuke, sname='Ashida']{Yosuke Ashida} 
\affiliation{Department of Physics, Tohoku University, Sendai, Miyagi 980-8578, Japan}
\email[]{yosuke.ashida.a1@tohoku.ac.jp}

\correspondingauthor{Ryo Sawada}

%%----------------------------------------------------------
%%                  ``abstract''
%%----------------------------------------------------------
%% Mark off the abstract in the ``abstract'' environment. 
\begin{abstract}
We investigate whether the Type Ibn supernova SN 2023uqf, reported close in time and direction to the $\sim$442 TeV IceCube alert IC-231004A, is physically consistent with a shock--circumstellar medium (CSM) interaction scenario. 
One-dimensional radiation-hydrodynamics calculations with {\tt STELLA} reproduce the ZTF optical light curves with a dense helium-rich CSM following $\rho_\mathrm{CSM} \propto r^{-3}$ and a CSM density parameter $D'\approx 50$. 
Using the shock evolution and CSM conditions obtained from the optical-fit RHD model,
we model time-dependent cosmic-ray acceleration and hadronic neutrino production during the interaction phase.
In our fiducial model, the inferred shock and CSM properties open a short-lived window in which multi-PeV hadron acceleration and efficient hadronic interactions can coexist.
Additional loss-limited effects could reduce the high-energy cutoff, but the model provides a useful optical-to-neutrino consistency test for a transient PeVatron-like phase.
After folding the predicted neutrino emission through the IceCube effective area, we obtain an expected number of $\sim10^{-5}-10^{-4}$ track-like events at $d = 723$ Mpc, depending on the alert selection. 
In the low-count regime, the model predicts a detection-time weighting for a rare event, and the detection time of IC-231004A falls within the high-weight interval while its energy scale is compatible with the modeled spectrum. 
Although a single event cannot establish a definitive association, our results show that the optically inferred environment of SN 2023uqf can satisfy the basic timing and energy requirements for a transient PeVatron-like phase in the fiducial model and illustrate how interaction-powered Type Ibn supernovae can be tested as high-energy neutrino sources.
\end{abstract}

%******************************************************************
%%----------------------------------------------------------
%%                  Introduction
%%----------------------------------------------------------

\section{Introduction}
High-energy neutrinos provide a direct probe of hadronic particle acceleration in astrophysical transients. 
Over the last decade, the astrophysical neutrino flux has been proven~\citep{2013PhRvL.111b1103A,2013Sci...342E...1I} and its spectrum is accurately measured by IceCube~\citep{2015PhRvD..91b2001A,2019PhRvD..99c2004A,2021PhRvD.104b2002A,2022ApJ...928...50A,2024PhRvD.110b2001A,2025PhRvD.112a2022A,2025arXiv250722234A,2025arXiv250722233A}.
In addition, there exist some successful observations of neutrinos in coincidence with a few confirmed sources, a blazar TXS~0506+056~\citep{2018Sci...361..147I} and a Seyfert galaxy NGC~1068~\citep{2022Sci...378..538I}, as well as possibly other Seyfert galaxies~\citep{2025arXiv251013403A}.
However, despite the growing list of individual source associations, population studies and stacking analyses indicate that the contributions from these identified classes fall short of accounting for the total diffuse neutrino flux measured by IceCube, suggesting that additional or as yet unidentified source populations must contribute significantly to the observed flux \citep{2021ApJ...921...45B}.

Interaction-powered core-collapse supernovae are a promising class in this context.
Recent theoretical studies have emphasized that the early ejecta--circumstellar medium (CSM) interaction phase can provide favorable conditions for diffusive shock acceleration up to (super-)PeV energies in interacting supernovae, aided by magnetic-field amplification and the transition to a collisionless shock \citep[e.g.,][]{2012IAUS..279..274K,2018PhRvD..97h1301M, 2025ApJ...984..103K,2026arXiv260206410E}. 

SN 2023uqf provides a rare opportunity to test this scenario in an event-specific way. 
The supernova was identified in optical surveys by the Zwicky Transient Facility (ZTF), and its sky position and explosion epoch were reported to be close to those of a high-energy neutrino alert detected by IceCube, IC-231004A \citep{2025arXiv250808355S}, whose reconstructed energy is \(E_\nu = 442.2\ {\rm TeV}\) \citep{2023GCN.34798....1B,2023GCN.34797....1I}. 
From an electromagnetic perspective alone, SN 2023uqf is classified as a Type Ibn supernova (SN Ibn) and exhibited bright and rapidly fading optical emission \citep{2025arXiv250808355S}. 
Although the exact peak of the optical light curve was not resolved, the source is confirmed to have reached an absolute \(g\)-band magnitude of at least \(M_g=-19.7\), corresponding to \(L\approx 2\times10^{43}\ {\rm erg\ s^{-1}}\). This places SN 2023uqf among the more luminous members of the Type Ibn class. 
In addition to its high luminosity, SN 2023uqf is also characterized by its extremely rapid fading. 
The half-peak timescale in the rest frame, \(t_{1/2}\), was only 8 days, making it the fastest-evolving event among the Type Ibn supernovae discovered by ZTF. 
Such rapid and luminous evolution indicates strong interaction between the supernova ejecta and a dense, helium-rich CSM, accompanied by a steep radial decline in the CSM density.

\citet{2025arXiv250808355S} presented the discovery, classification, multi-wavelength follow-up, and chance-coincidence assessment of SN 2023uqf, and modeled its optical light curve with the semi-analytic CSM-interaction framework implemented in MOSFiT. 
Their analysis provides the observational basis for the present study, including the Type Ibn classification and multi-wavelength constraints.
However, a semi-analytic, single-zone light-curve fit is not designed to follow the hydrodynamic shock evolution directly, and therefore does not uniquely specify the dynamical shock history required for particle-acceleration calculations, including the time-dependent forward-shock radius, shock velocity, CSM density encountered by the shock, and the epoch at which the shock can become collisionless. 
These quantities directly control the maximum cosmic-ray energy and the hadronic neutrino yield.

Building on these observational constraints, we move from a phenomenological light-curve characterization to an event-specific radiation-hydrodynamic model of SN 2023uqf, and use the resulting shock--CSM evolution directly in a time-dependent neutrino calculation. 
Our aim is not to revise the chance-coincidence estimate of \citet{2025arXiv250808355S}, but to test whether the optically inferred interaction environment can physically support a transient PeVatron phase consistent with the timing and energy scale of IC-231004A. 
This provides an optical-to-neutrino dynamical consistency test that was not the focus of \citet{2025arXiv250808355S}.

In this Letter, we first construct one-dimensional radiation-hydrodynamic light-curve models for SN 2023uqf using Type Ibn-motivated ejecta and CSM structures, and infer the shock radius, shock velocity, and CSM density relevant for non-thermal particle acceleration (Section \ref{sec:mm_em}). We then calculate the high-energy neutrino emission expected from this model (Section \ref{sec:mm_hinu}). 
Finally, based on IceCube's effective area, we verify whether this is consistent with the IC-231004A event in terms of detection efficiency and temporal consistency  (Section \ref{sec:mm_nuobs}). 
We do not claim that a single neutrino establishes a definitive association. Rather, we show that the optical properties of SN 2023uqf imply a short-lived shock-interaction phase in which the energy and timing requirements for a \(\sim 442\) TeV neutrino can be physically tested.

%******************************************************************
%----------------------------------------------------
%                   Model 	
%----------------------------------------------------
%\section{Physical properties of SN 2023\lowercase{uqf} inferred from Optical emission} \label{sec:mm_em}

\begin{figure}[htb]
    \centering
    \includegraphics[width=0.45\textwidth]{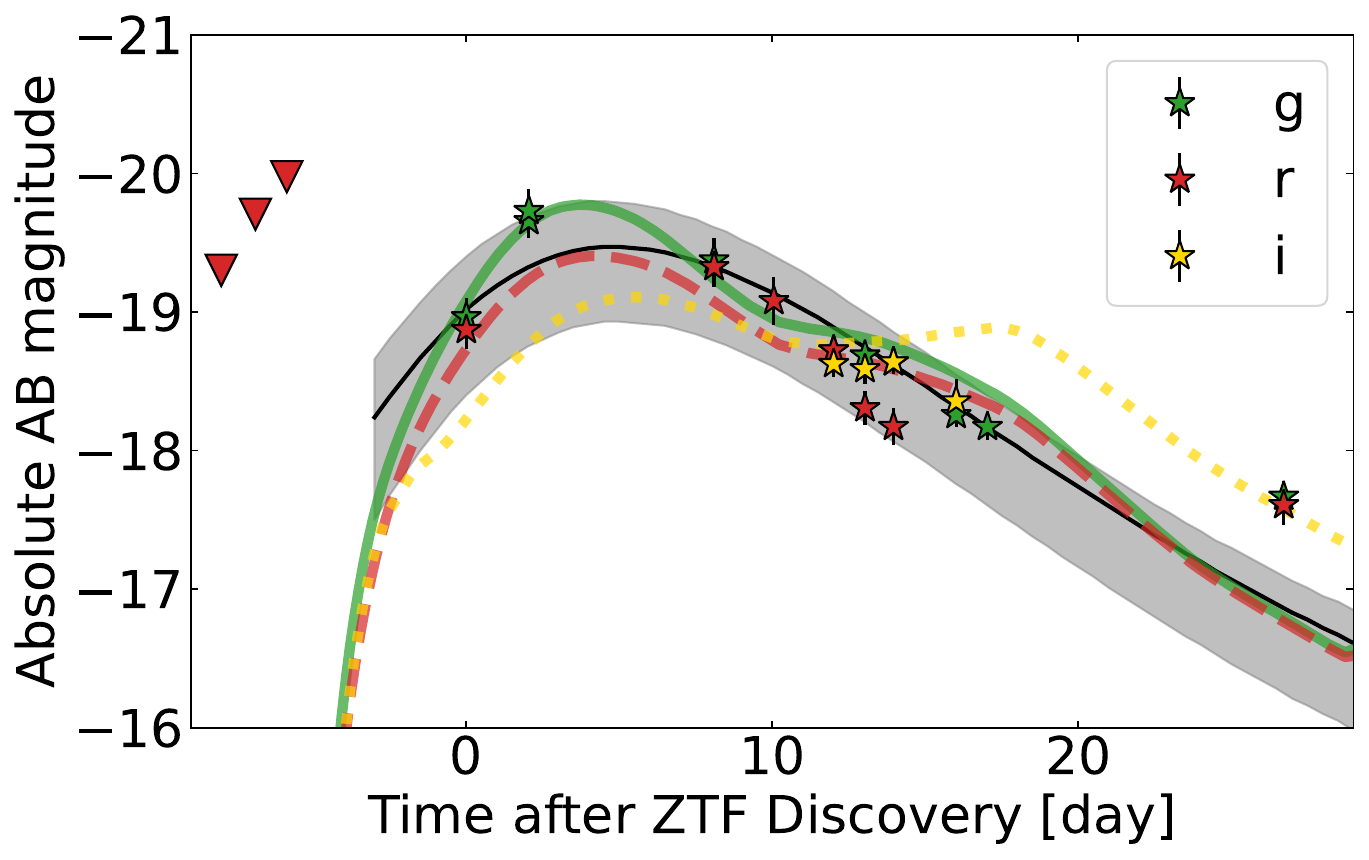}
    \caption{Observed and synthesized optical light curves are shown. 
    The shown points of $g$(green), $r$(red) and $i$(yellow)- bands are taken from \citet{2025arXiv250808355S}.
    Star symbols mean detections, and downward triangles mean upper limits from non-detections.
    The solid green line, dashed red line, and dotted yellow line correspond to synthetic $g$, $r$ and $i$- bands light curves, respectively.
    The synthesized light curves are shifted by $\Delta t=-10$ days relative to the ``time after ZTF discovery'' axis, corresponding to an inferred explosion epoch $\sim$10 days prior to the first ZTF detection. 
    Gray regions correspond to the average light curve and 1.96$\sigma$ error bars of 18 Type Ibn SNe from \citet{2017ApJ...836..158H}. The average light curve is displayed with the x-axis shifted to align the peak positions.}
    \label{fig:em_LC}
\end{figure}

\section{Optical modeling of SN 2023\lowercase{uqf} } \label{sec:mm_em}

The goal of this section is to construct a physically constrained radiation-hydrodynamic model of the optical emission of SN 2023uqf that can also provide the dynamical quantities required for the neutrino calculation.
Figure \ref{fig:em_LC} shows the resulting comparison between the observed ZTF \textit{g, r}, and \textit{i}-band light curves and our synthetic light curves.
The model is not intended as a unique inversion of the optical data, but as a realization of the shock–CSM interaction under Type Ibn-motivated CSM assumptions.
The dynamical evolutions are then used as inputs to the particle-acceleration and neutrino-production calculations in Section \ref{sec:mm_hinu}.

\subsection{Radiation-hydrodynamic setup} \label{subsec:init_opt}

We calculate the optical light curves with the one-dimensional, frequency-dependent radiation-hydrodynamics code {\tt STELLA} \citep{1998ApJ...496..454B,2000ApJ...532.1132B,2006A&A...453..229B}, which follows the conversion of shock-deposited energy into emergent radiation.

We adopt a representative ejecta model motivated by previous Type Ibn progenitor and light-curve studies. 
We therefore fix the ejecta mass and kinetic energy to \(M_{\rm ej}=2\,M_\odot\) and \(E_{\rm expl}=10^{51}\ {\rm erg}\), consistent with optical light-curve modeling of typical SNe Ibn \citep[e.g.,][]{2016ApJ...824..100M,2022ApJ...927...25M}.
This choice reduces the degeneracy between ejecta and CSM parameters and allows us to focus on the CSM density structure, which is the quantity most directly relevant for both the optical luminosity and hadronic neutrino production.

We initialize the interaction calculation at $R_\mathrm{sh,0}=10^{13}\,\mathrm{cm}$. 
We emphasize that this choice is not an additional parameter tuned to the optical light curve.
Given the compact helium-star (WR-like; $R_\star\sim10^{11}\,\mathrm{cm}$) progenitors generally inferred for SNe Ibn \citep[e.g.,][]{2017ApJ...836..158H,2025arXiv251112362F}, this radius is already much larger than the progenitor radius \((R_{\rm sh,0}\gg R_\star)\), so the ejecta are expected to be already close to homologous expansion. 
The corresponding travel time is
\begin{equation}
R_{\rm sh,0}/v_{\rm ej}
    \simeq 0.1
    \left(\frac{R_{\rm sh,0}}{10^{13}\ {\rm cm}}\right)
    \left(\frac{v_{\rm ej}}{1.0\times10^9\ {\rm cm\ s^{-1}}}\right)^{-1}
    {\rm day},    
\end{equation}
which is much shorter than the observed optical evolution; therefore, starting the interaction calculation at \(R_{\rm sh,0}\) does not affect the day-scale light-curve and neutrino calculations.

For the ejecta profile, we adopt the standard homologously expanding broken power-law structure widely used in supernova interaction calculations \citep[e.g.,][]{1999ApJ...510..379M,2006ApJ...651..381C}.
This prescription is not tuned to SN 2023uqf, but provides a conventional compact-progenitor ejecta structure on which the CSM interaction is calculated. 
Specifically, the ejecta density is specified in velocity space as
\begin{equation}
    \rho_{\rm ej}(v,t)=\rho_{\rm t}(t)
    \begin{cases}
    1, & v < v_{\rm t},\\
    (v/v_{\rm t})^{-n}, & v \geq v_{\rm t},
    \end{cases}
\end{equation}
where $v=r/t$, $n=7$, and the inner profile is flat $(\delta=0)$. 
The transition velocity is set by the adopted ejecta mass and kinetic energy,
\begin{equation}
v_{\rm t} =
\left[
\frac{2(5-\delta)(n-5)}{(3-\delta)(n-3)}
\frac{E_{\rm expl}}{M_{\rm ej}}
\right]^{1/2},    
\end{equation}
which gives $v_{\rm t}\simeq 6.5\times10^8\ {\rm cm\ s^{-1}}$ for our fiducial parameters. 
Thus, the boundary between the inner and the outer ejecta is not a fixed radius; it is located at $r_{\rm t}(t)=v_{\rm t}t$ in the homologously expanding ejecta.

For the CSM, we adopt a power-law density profile,
\begin{equation}
    \rho_{\rm CSM}(r)=D r^{-k},
\end{equation}
where \(k\) is the radial density slope. 
We use \(k=3\) as our fiducial Type Ibn-motivated model and write the corresponding normalization as
\begin{equation}
    \rho_\mathrm{CSM}(r) 
    = 10^{-14}D' \left( \dfrac{r}{5\times10^{14}\,\mathrm{cm}} \right)^{-3} \,\mathrm{g\,cm^{-3}}~,
\end{equation}
where $D' = 1$ 
provides a reference normalization inferred in previous Type Ibn optical and X-ray modeling
\citep{2022ApJ...927...25M,2025ApJ...980...86I}.
We assume a helium-rich CSM composition with $(X_\mathrm{He}, X_\mathrm{C}, X_\mathrm{O})=(0.9, 0.06, 0.04)$, following \citet{2008MNRAS.389..141M} and \citet{2009MNRAS.400..866C} \citep[see also, e.g.,][for quantitative discussion on CSM abundance of SNe Ibn]{2009ApJ...692..546S,2022A&A...658A.130D,2025ApJ...980...86I}.
The choice of \(k=3\) is motivated by previous Type Ibn light-curve modeling, which favors a compact, steeply declining CSM over an extended wind-like \(\rho\propto r^{-2}\) profile for rapidly fading SNe Ibn \citep{2022ApJ...927...25M}. 
A wind-like profile tends to produce a slower late-time decline unless the rapid fading is imposed mainly by the CSM outer edge. 
We therefore adopt \(k=3\) as a fiducial steep CSM profile for the luminous and rapidly fading SN 2023uqf.

Figure \ref{fig:em_LC} compares the observed and synthetic $g,r,$ and $i$-band light curves of SN 2023uqf. 
With these physically motivated choices fixed, a dense CSM model with \(D'=50\) matches the observed multi-band evolution up to \(\approx30\) days after ZTF discovery.
Such a high CSM density is consistent with the high optical luminosity of SN 2023uqf, which lies on the luminous side of the Type Ibn population \citep[Figure \ref{fig:em_LC};][]{2017ApJ...836..158H}.

This value should not be interpreted as a unique inversion of the CSM density normalization, because the ejecta mass and kinetic energy are fixed in our fiducial RHD setup. 
Rather, the optical light curve constrains the shock-dissipation history, \(L_{\rm sh}\sim2\pi R_{\rm sh}^2\rho_{\rm CSM}V_{\rm sh}^3\), which is also the quantity that normalizes the neutrino calculation.
In this sense, our RHD interpretation places SN 2023uqf on the dense-CSM side of the Type Ibn population. 
At the same time, Figure~\ref{fig:em_LC} shows that SN 2023uqf is not an obvious photometric outlier among Type Ibn events.
This expected CSM density $(D'=50)$ gives
\begin{equation}
   \rho_{\rm CSM}(10^{13}\ {\rm cm}) \simeq 6.3\times10^{-8}\ {\rm g\ cm^{-3}}, 
\end{equation}
at the inner radius of the helium CSM. 
At such small radii, the optical depth of the adopted CSM is still large, so the early shock is expected to remain radiation-mediated.
We therefore do not assume non-thermal acceleration from \(R_{\rm sh,0}\); instead, the acceleration onset is determined separately by the breakout/collisionless-shock condition in Section \ref{subsec:cr}.

The remaining band-dependent residuals at \(\gtrsim10\) days, including the slight overprediction after \(\sim10\) days and the underprediction of the \(g\)- and \(r\)-band fluxes around \(\sim30\) days, likely reflect the simplified single power-law CSM structure and the approximate treatment of the late-time color and thermal evolution. These late-time residuals do not affect the early post-breakout shock phase that dominates the neutrino output in the fiducial calculation.

The ZTF photometry compiled by \citet{2025arXiv250808355S} is shown as a function of time after the first ZTF detection (``discovery''), while the true explosion epoch is not directly measured. 
Since our shock–CSM interaction models are computed as a function of time since explosion, we treat the explosion time $t_\mathrm{expl}$ as a parameter and allow for a rigid time translation between the model and the data. 
We find that shifting the model light curves by $\Delta t=-10$ days relative to the ZTF-discovery time axis provides a reasonable description of the observed light-curve evolution, implying $t_\mathrm{expl}\approx t_\mathrm{disc}-10\,\mathrm{days}$. 
This value differs from the MOSFiT explosion epoch reported by \citet{2025arXiv250808355S}, but the two should not be interpreted as direct measurements of the same quantity. 
The MOSFiT value is obtained within a semi-analytic light-curve prescription, whereas our time shift is tied to the RHD evolution of a Type Ibn CSM-interaction model.
We therefore adopted this inferred explosion epoch when placing the optical and neutrino evolution on a common timeline.

\subsection{Shock evolution from the RHD model}

\begin{figure}[tb]
    \centering
    \includegraphics[width=0.45\textwidth]{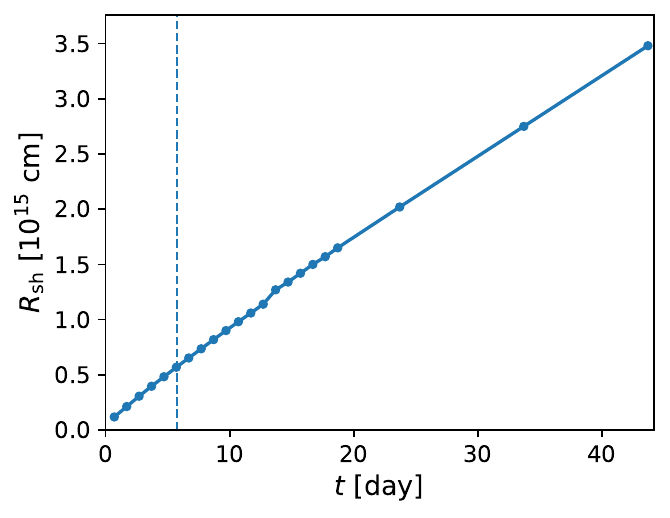}
    \includegraphics[width=0.45\textwidth]{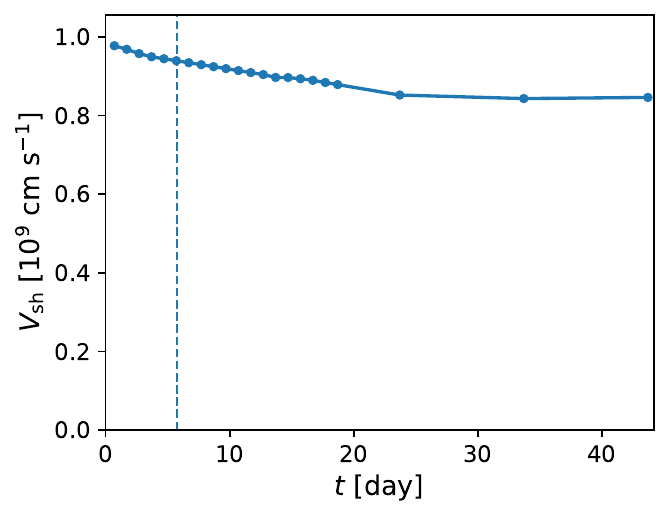}
    \includegraphics[width=0.45\textwidth]{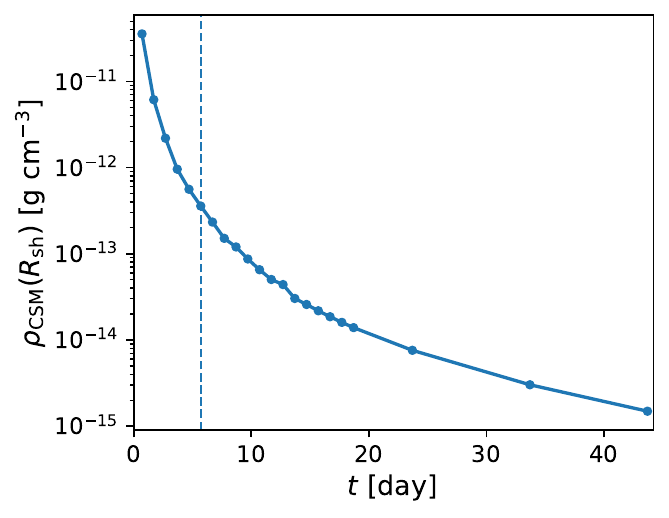}
    \caption{RHD-derived shock evolution used as input to the neutrino calculation.
    Top: forward-shock radius \(R_{\rm sh}(t)\). 
    Middle: shock velocity \(V_{\rm sh}(t)\). 
    Bottom: CSM density encountered by the shock,\(\rho_{\rm CSM}(R_{\rm sh}(t))\). 
    The vertical dashed line marks \(t_{\rm onset}=5.74\) days, after which collisionless acceleration is allowed in the neutrino calculation.}
    \label{fig:hydro_evolution}
\end{figure}

The RHD calculation provides the dynamical shock quantities used in the neutrino calculation.
Figure~\ref{fig:hydro_evolution} summarizes these outputs: the forward-shock radius \(R_{\rm sh}(t)\), shock velocity \(V_{\rm sh}(t)\), and CSM density encountered by the shock, \(\rho_{\rm CSM}(R_{\rm sh}(t))\).
We emphasize that these quantities are outputs of the same model that reproduces the optical light curves in Figure~\ref{fig:em_LC}; they are not adjusted independently in the neutrino calculation.

Figure \ref{fig:hydro_evolution} shows that the forward shock expands almost linearly with time. This reflects the fact that the shock velocity remains nearly constant at \(V_{\rm sh}\sim10^9\ {\rm cm\ s^{-1}}\) over the relevant phase, so that the shock radius approximately follows \(R_{\rm sh}\simeq V_{\rm sh}t\). 
Thus, the RHD model confirms a nearly free-expansion shock evolution during the phase used for the neutrino calculation. Meanwhile, the CSM density encountered by the shock decreases rapidly with time because the shock propagates through the steep \(k=3\) density profile.
The vertical line marks \(t_{\rm onset}=5.74\) days, after which collisionless acceleration is allowed according to the breakout condition discussed in more detail in Section \ref{subsec:cr}.

\subsection{Consistency with X-ray and radio non-detections}\label{subsec:multi_em}

Interaction with dense CSM is generally expected to produce luminous X-ray and radio emission in addition to the optical output
\citep[e.g.,][for a review]{2017hsn..book..875C,2025Univ...11..363C}.
For SN 2023uqf, however, \citet{2025arXiv250808355S} reported non-detections in a
soft X-ray band (0.3--10 keV; 24 days after ZTF discovery) and a radio band
(10 GHz; 20 and 75 days after ZTF discovery), with upper limits of
$L_X^{\rm lim}=7\times10^{41}~{\rm erg\,s^{-1}}$ and
$L_{\rm radio}^{\rm lim}=10^{38}~{\rm erg\,s^{-1}}$, respectively.
Here we show that the dense \(k=3\) CSM inferred from the optical model is consistent with these limits, basically following \citet{2017hsn..book..875C}. 
For this estimate, we adopt $R_{\rm sh}=V_{\rm sh}t$ with $V_{\rm sh}\simeq10^{9}~{\rm cm\,s^{-1}}$ (see also Fig. \ref{fig:hydro_evolution}).

For the radio band, free--free absorption by the ionized helium-rich CSM gives
\begin{equation}
    \tau_{\rm ff}(\nu)
    \simeq
    8\times10^{11}
    \left(\frac{T_e}{10^5\,{\rm K}}\right)^{-3/2}
    \left(\frac{\nu}{10\,{\rm GHz}}\right)^{-2}
    \left(\frac{t}{\rm day}\right)^{-5}.    
    \label{eq:tauff}
\end{equation}
At the two epochs of the 10 GHz observations, \(t\simeq30\) and \(85\) days at explosion-frame, Eq. \eqref{eq:tauff} gives \(\tau_{\rm ff}\simeq3.3\times10^4\) and
\(\simeq1.8\times10^2\), respectively, for the least-absorbing case $T_{e}=10^{5}$ K \citep{1998ApJ...509..861F}.
The model therefore predicts strong radio absorption at the observed epochs, naturally explaining the radio non-detections.

For soft X-rays, the dominant opacity is photoelectric absorption, mainly by the K-shell electrons of C/O in the helium-rich CSM \citep[see also][]{2025ApJ...980...86I}.
Using the bound-free opacity scaling
\begin{equation}
\kappa_{\rm bf}(E)\simeq 1.06\times10^{2}
\left(\frac{E}{\rm keV}\right)^{-8/3}\frac{Z}{Z_{\odot}}~{\rm cm^{2}\,g^{-1}},
\label{eq:kappabf}
\end{equation}
with \(Z/Z_\odot\simeq7\), and the unshocked CSM column \(\Sigma(t)=D/2R_{\rm sh}^2\), 
we obtain the optical depth
\begin{equation}
    \tau_{\rm bf}(E,t)
    \simeq
    3.2\times10^6
    \left(\frac{E}{\rm keV}\right)^{-8/3}
    \left(\frac{t}{\rm day}\right)^{-2}.    
\end{equation}
At the epoch of the X-ray upper limit, \(t\simeq34\) days after explosion, this gives \(\tau_{\rm bf}\simeq2.7\times10^3\) at 1 keV and \(\simeq6\) at 10 keV.
While the soft X-ray band is clearly strongly absorbed, the optical depth becomes substantially smaller at the hard end of the observed band. We therefore estimate the intrinsic X-ray luminosity to check whether the remaining absorption is sufficient to suppress the escaping flux below the observational upper limit.

Assuming electron-ion temperature equilibration behind the forward shock, the post-shock temperature for \(V_{\rm sh}\simeq10^9\ {\rm cm\ s^{-1}}\) is \(k_{\rm B}T_{\rm FS}\simeq260\) keV. 
A one-zone free--free estimate for a shocked shell with fractional thickness \(f=0.1\) gives a bolometric luminosity \(L_{\rm ff}^{\rm bol}\simeq4.4\times10^{43}\ {\rm erg\ s^{-1}}\) at the epoch of the X-ray upper limit. 
Since the spectrum is hard, only a fraction \(f_{0.3-10\,{\rm keV}}\simeq0.05\) falls in the observed 0.3--10 keV band, yielding 
\begin{equation}
    L_X^{\rm int}(0.3\text{--}10\,{\rm keV})
    \simeq 2\times 10^{42}\ {\rm erg\ s^{-1}}
    \simeq 3 L_X^{\rm lim}.
\end{equation}
Thus, an effective optical depth of only \(\tau_{\rm eff}\gtrsim\ln(L_X^{\rm int}/L_X^{\rm lim})\sim1\) is sufficient to suppress the escaping X-ray flux below the observed limit. 
This requirement is satisfied by the bound-free opacity above: the soft part of the band is completely absorbed, and even at 10 keV we find \(\tau_{\rm bf}\simeq6\) at the X-ray epoch.

We therefore conclude that the X-ray and radio non-detections are consistent with the dense CSM model, although a full treatment of photoionization and radiative transfer will be required for more precise multi-wavelength predictions.

\section{High-energy Neutrino Emissions} \label{sec:mm_hinu}
We now evaluate the time-dependent high-energy neutrino emission expected from SN 2023uqf under the dense-CSM scenario constrained by the optical light curves in Section \ref{sec:mm_em}. 
The neutrino calculation is based on the same initial shock position and subsequent evolution as in Section \ref{sec:mm_em}.
We adopt the forward-shock radius $R_\mathrm{sh}(t)$, shock velocity $V_\mathrm{sh}(t)$, and CSM density encountered by the shock $\rho_\mathrm{CSM}(R_\mathrm{sh})$ as derived from the radiation-hydrodynamic light-curve modeling. 
We therefore construct a minimal, physically motivated model of shock-driven cosmic-ray (CR) acceleration and hadronic neutrino production during the shock–CSM interaction phase, following established time-dependent frameworks \citep{2018PhRvD..97h1301M,2024PhRvD.109j3020M,2025ApJ...984..103K,2025ApJ...982...93S} and extending them to include a helium-primary composition.

\subsection{Shock-driven CR acceleration}\label{subsec:cr}
A key uncertainty in early-time supernova particle acceleration is whether the forward shock is radiation-mediated or collisionless. Radiation-mediated shocks suppress diffusive shock acceleration, while collisionless shocks allow efficient non-thermal particle acceleration. Following \citet{2018PhRvD..97h1301M}, we approximate the onset of collisionless acceleration by the photon breakout time, $t_\mathrm{onset} \approx t_\mathrm{bo}$, defined by the Thomson optical-depth condition ahead of the forward shock,
\begin{align} 
    \tau_\mathrm{T} &= 
    \frac{\sigma_\mathrm{T}}{\mu_e m_\mathrm{p}} 
    \int_{R_\mathrm{sh}}^\infty Dr^{-3}dr \nonumber \\
    &=\frac{\sigma_\mathrm{T}D}{2\mu_e m_\mathrm{p}}R_\mathrm{sh}^{-2}\lesssim \frac{c}{V_\mathrm{sh}}~, \label{eq:onset}
\end{align}
where $\sigma_\mathrm{T}$ is the Thomson cross section, $\mu_e$ is the mean molecular weight per electron of the unshocked CSM ($\mu_e=2$ for fully ionized helium), and $V_\mathrm{sh}$ is the velocity of the forward shock propagating in the CSM.
For the fiducial \(D'=50\) model, this condition gives the approximate onset time of collisionless acceleration
\begin{equation}
    t_{\rm onset}\sim 5.3 \left(\frac{D'}{50}\right)^{1/2} \left(\frac{V_{\rm sh}}{10^9\ {\rm cm\ s^{-1}}}\right)^{-1/2} {\rm day}
\end{equation}
for a characteristic shock velocity of \(10^9\ {\rm cm\ s^{-1}}\).
Using the RHD shock trajectory, we obtain \(t_{\rm onset}\simeq 5.74\) days after core collapse.
This time defines the beginning of the CR-acceleration and neutrino calculation below.

Once collisionless acceleration sets in, we assume that CRs are accelerated at the forward shock with a power-law spectrum with an exponential cutoff. 
To allow for nuclear primaries, we write the per-nucleus spectrum as
\begin{equation}
    \frac{dn_\mathrm{CR}^{(A)}}{dE_A} \propto E_A^{-s}\exp{\left( -\dfrac{E_A}{E^\mathrm{max}_A} \right)}~,
\end{equation}
where $A$ and $Z$ are the mass and charge numbers of the primary nucleus.
As a fiducial composition, we take the accelerated CR nuclei to be helium, \(A=4\) and \(Z=2\), motivated by the He-rich CSM inferred for SNe Ibn in Section \ref{subsec:init_opt}. 
This should be regarded as an idealized nuclear-primary case, not as an additional parameter tuned to SN 2023uqf.

The CR spectrum is normalized by the CR energy density,
\begin{equation}
    u_\mathrm{CR}
    =\int dE \, E\frac{dn_\mathrm{CR}}{dE}
    = \epsilon_{p} \cfrac{\rho_\mathrm{CSM} V_\mathrm{sh}^2}{2}~.     
\end{equation}
Here, we adopt $s = 2$ and $\epsilon_{p}=0.1$ as fiducial values of efficient shock acceleration \citep{2014ApJ...783...91C}.

The maximum CR energy is determined by requiring the acceleration time to be shorter than the characteristic escape time $t_\mathrm{acc}(E^\mathrm{max}_A)\leqq t_\mathrm{esc}$. 
Using the standard diffusive shock-acceleration estimate \citep{1983RPPh...46..973D},
\begin{equation}
    t_\mathrm{acc}(E_A) \approx \frac{20}{3}\frac{cE_A}{ZeBV_\mathrm{sh}^2}
    \lesssim \frac{R_\mathrm{sh}}{c}~,
\end{equation}
we compute $E^\mathrm{max}_A$ as a function of the position of the forward shock, which corresponds to time.
We parameterize the magnetic-field strength through a microphysical fraction $\epsilon_B$ and adopt $\epsilon_B=0.01$ as a fiducial value. 
For reference, the above condition gives the escape-limited maximum energy
\begin{align}
    E_{A}^{\max}
    \simeq 8.9\,Z
    &\left(\frac{\epsilon_B}{0.01}\right)^{1/2}
    \left(\frac{\rho_{\rm CSM}}{10^{-12}\ {\rm g\ cm^{-3}}}\right)^{1/2} \nonumber \\
    &\left(\frac{R_{\rm sh}}{5\times10^{14}\ {\rm cm}}\right)
    \left(\frac{V_{\rm sh}}{10^9\ {\rm cm\ s^{-1}}}\right)^3
    {\rm PeV}.    
\end{align}
For the adopted CSM profile $(k=3$ and $D'=50)$, the same estimate becomes
\begin{align}
    E_{A}^{\max}
    \simeq
    6.3\,Z
    &\left(\frac{\epsilon_B}{0.01}\right)^{1/2}
    \left(\frac{D'}{50}\right)^{1/2}
    \nonumber \\
    &\left(\frac{R_{\rm sh}}{5\times10^{14}\ {\rm cm}}\right)^{-1/2}
    \left(\frac{V_{\rm sh}}{10^9\ {\rm cm\ s^{-1}}}\right)^3
    {\rm PeV}.   
\end{align}
Compared with a proton-primary model at the same rigidity-limited acceleration condition, the helium-primary case has a larger total nuclear cutoff by a factor \(Z=2\), but a lower cutoff per nucleon by \(Z/A=1/2\).

The cutoff adopted in the fiducial neutrino calculation is therefore an escape-limited, rather than fully loss-limited, maximum energy. 
We do not explicitly impose additional loss limits from photomeson production, Bethe--Heitler pair production, photodisintegration, or inelastic nuclear collisions in the dense CSM. 
These processes can reduce \(E_A^{\max}\) and will be addressed in future work.
The spectra shown below should therefore be interpreted as escape-limited reference spectra; a fully loss-limited treatment would primarily shift the high-energy cutoff to lower energies.

\subsection{Hadronic neutrino production and transport}
Neutrinos are produced when shock-accelerated CRs interact hadronically with the dense CSM, dominantly through pion production and decay (e.g., $pp \rightarrow \pi^{+} \rightarrow \mu^{+}\nu_{\mu} \rightarrow e^{+}\nu_{e}\nu_{\mu}\bar{\nu}_{\mu}$ and charge conjugates). 
In the fiducial model, this corresponds to interactions of
helium-primary CR nuclei with a helium-rich CSM. We treat these
nuclear interactions using the superposition approximation: a nucleus
with mass number \(A\) and energy \(E_A\) is represented as \(A\)
nucleons with per-nucleon energy \(E_p=E_A/A\), interacting with
the target nucleon density \(n_{\rm N}\simeq\rho_{\rm CSM}/m_p\).
Differences between \(pp\), \(pn\), and \(nn\) channels are neglected.
The differential production rate $ \Phi(E_\nu) dE_\nu$ in the energy interval $(E_\nu, E_\nu+ dE_\nu)$ per unit nucleon number density is
\begin{equation}
    \Phi(E_\nu) \approx 
    \int_{E_\mathrm{min}}^{\infty} 
    \sigma_{pp}\left(E_p\right)\cdot  
    F_\nu\left(E_\nu, E_p\right) \cdot 
    A\cdot n^{(A)}_\mathrm{CR}\left(E_A\right) \frac{dE_p}{E_p} ~,
\end{equation}
where $F_\nu(E_\nu, E_p)$ is the differential spectrum of the secondary particles, and its parametrization follows Eqs. (62) and (66) of \citet{2006PhRvD..74c4018K}.  
The function $\sigma_{pp}(E_p)$ is the inelastic cross section of $pp$ interactions, following the post-Large-Hadron-Collider formula given by \citet{2014PhRvD..90l3014K}.

The time-dependent neutrino injection rate [s$^{-1}$ TeV$^{-1}$ cm$^{-3}$] is then
\begin{equation}
    \dot{n}^\mathrm{inj}_\nu(E_\nu,t) = c\,\frac{\rho_\mathrm{CSM}(R_\mathrm{sh}(t))}{m_\mathrm{p}} \cdot \Phi(E_\nu,t)~,
\end{equation}
where \(\rho_{\rm CSM}/m_p\) is the target nucleon density in the helium-rich CSM under the same superposition approximation. We then solve the transport equation
\begin{equation}
    \frac{\partial}{\partial t} n_\nu(E_\nu,t)= - \frac{n_\nu(E_\nu,t)}{t_\mathrm{esc}}+ \dot{n}^\mathrm{inj}_\nu(E_\nu,t)~,
\end{equation}
to obtain the escaping neutrino energy luminosity [erg s$^{-1}$],
\begin{equation}
    L_\nu(E_\nu,t) = \frac{E_\nu^2 \cdot n_{\nu}(E_\nu,t) \cdot \mathcal{V}_\mathrm{em}(t)}{t_\mathrm{esc}}~,
\end{equation}
where $\mathcal{V}_\mathrm{em}(t)\approx(4\pi/3)R_\mathrm{sh}^3$ is the volume of the emission region.
We also define the corresponding source-frame differential neutrino number luminosity $\Phi_{\rm SN\mathchar`-\nu}(E_\nu,t)$,
\begin{equation}
\Phi_{\rm SN\mathchar`-\nu}(E_\nu,t)
\equiv
\frac{n_\nu(E_\nu,t)\,\mathcal{V}_{\rm em}}{t_{\rm esc}},\label{eq:Ndotnu}
\end{equation}
which has units of [${\rm s^{-1}\,TeV^{-1}}$].
This setup yields a self-consistent, time-dependent prediction for $L_\nu(E_\nu,t)$ and $\Phi_{\rm SN\mathchar`-\nu}(E_\nu,t)$ given the shock evolution and CSM properties inferred from the optical data.

\begin{figure}[tb]
    \centering
    \includegraphics[width=0.45\textwidth]{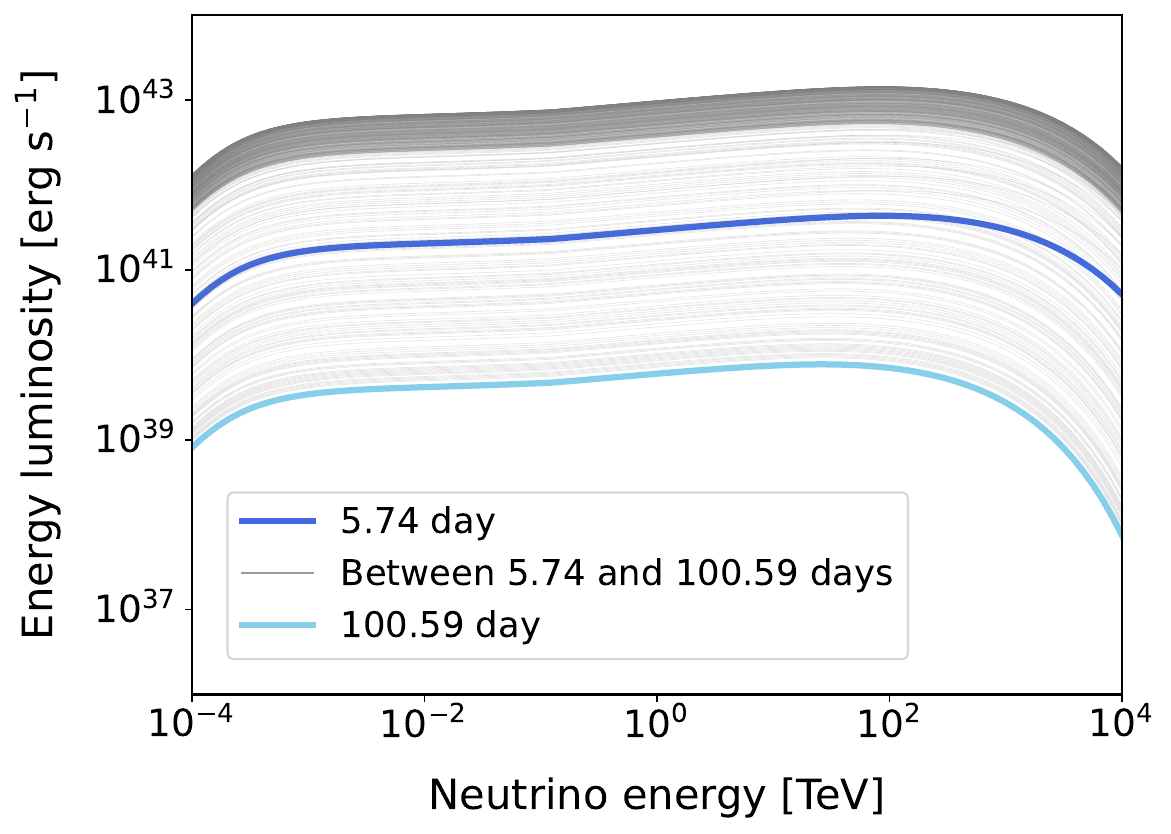}\vspace{+1mm}
    \includegraphics[width=0.43\textwidth]{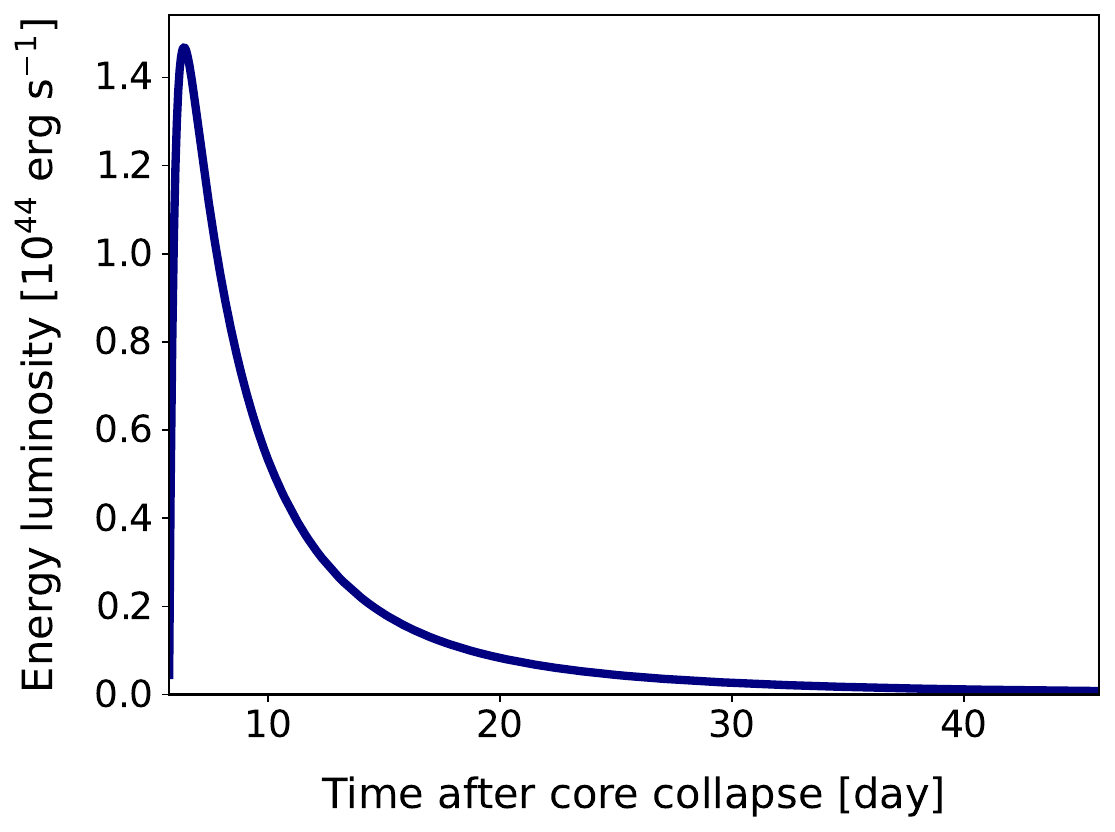}
    \caption{Top: the neutrino spectral energy luminosity at different time slices, with the blue lines for the calculation boundaries and the gray ones for in-between times. 
    The blue curves denote the onset and end of the neutrino calculation, \(t=5.74\) and \(100.59\) days after core collapse, respectively, while the gray curves show intermediate epochs.
    Bottom: the total energy luminosity of SN neutrinos as a function of days from the core collapse. Here, the sum of six neutrino flavors ($\nu_{e}+\bar{\nu}_e+\nu_{\mu}+\bar{\nu}_\mu+\nu_{\tau}+\bar{\nu}_\tau$) is shown in both panels.}
    \label{fig:elumi}
\end{figure}

Figure \ref{fig:elumi} presents the resulting neutrino spectra and bolometric neutrino luminosity evolution expected from the shock-CSM interaction phase under the optical-fit CSM conditions. 
The top panel shows the neutrino spectral energy luminosity at selected epochs, illustrating how the spectral normalization and cutoff evolve as the shock propagates and the maximum CR energy changes. 
The bottom panel shows the corresponding neutrino luminosity as a function of time (summed over flavors). 
We fold this time-dependent emission with the IceCube effective area to obtain the expected detection rate and its time dependence in Section \ref{sec:mm_nuobs}, enabling a direct comparison with the reported $\sim$442 TeV event IC-231004A.
In the low-event count relevant to SN 2023uqf, this time dependence primarily determines when a rare detection would be most likely.

We note that cosmological time dilation and redshift would rescale observer-frame times and energies by $(1+z)$. For $d=723$ Mpc, which corresponds to $z\sim0.1-0.2$, this is at the $\lesssim20\%$ level and does not affect our qualitative conclusions, so we ignore it for simplicity.

Under the optical-fit shock and CSM conditions, the source enters a narrow regime in which multi-PeV hadron acceleration is feasible while hadronic interactions remain efficient. In SN 2023uqf, this overlap is enabled by the combination of a compact progenitor, which gives a rapidly expanding early shock, and a steep dense CSM profile $(\rho_\mathrm{CSM}\propto r^{-3})$, which keeps the target density high at small radii.

\section{High-energy Neutrino Detection} \label{sec:mm_nuobs}

We now convert the source-frame neutrino emission predicted in Figure \ref{fig:elumi} into an observational expectation for IceCube.
We refer to the effective area released from IceCube, particularly in its catalog ``IceCat-1''~\citep{2023ApJS..269...25A}, for the track-like events induced by muon neutrinos.
The referred event, IC-231004A, was released as having come from the declination angle of 25.04 deg, hence, in this study, the IceCat-1 effective areas corresponding to the declination angles of 0 to 30~deg are used. 
The expected event rate at IceCube from a neutrino source is written as 
\begin{equation}
    \frac{dN_{\rm event}}{dt} = \frac{1}{4 \pi d^2} \int_{E_{\rm min}}^{E_{\rm max}} \frac{\Phi_{\rm SN\mathchar`-\nu}(E_\nu,t)}{3} \cdot \mathcal{A}_{\rm eff}(E_\nu) dE_\nu~, \label{eq:dNdt}
\end{equation}
where $\Phi_{\rm SN\mathchar`-\nu}(E_\nu,t)$ is the source-frame differential neutrino number luminosity [s$^{-1}$ TeV$^{-1}$], $\mathcal{A}_{\rm eff}(E_\nu)$ is the effective area, and $d$ is distance to the source. 
We take $d = 723$~Mpc for SN~2023uqf and the analysis energy range of $E_{\rm min} = 10$~TeV and $E_{\rm max} = 10$~PeV.
Here, the flux is divided by 3 to estimate only for muon-type neutrinos ($\nu_\mu + \bar{\nu}_\mu$).

\begin{figure}
    \centering
    \includegraphics[width=0.45\textwidth]{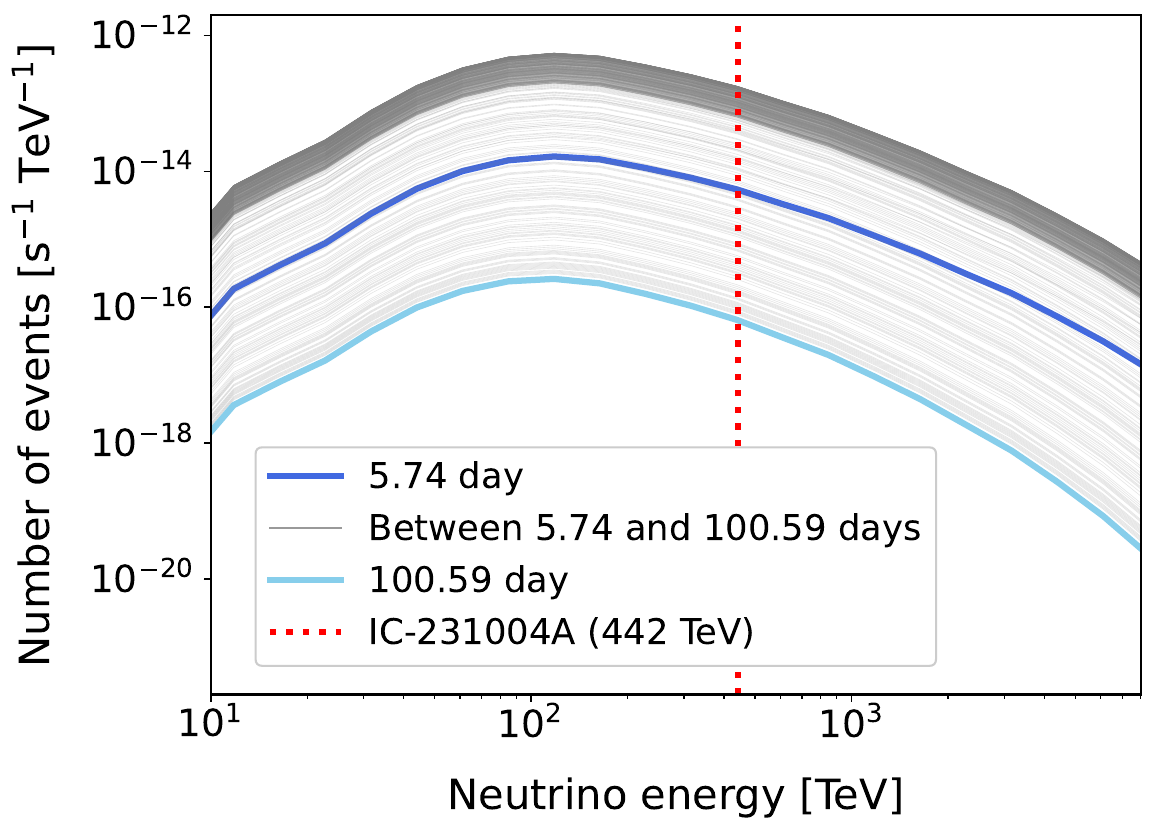}\vspace{+3mm}
    \includegraphics[width=0.48\textwidth]{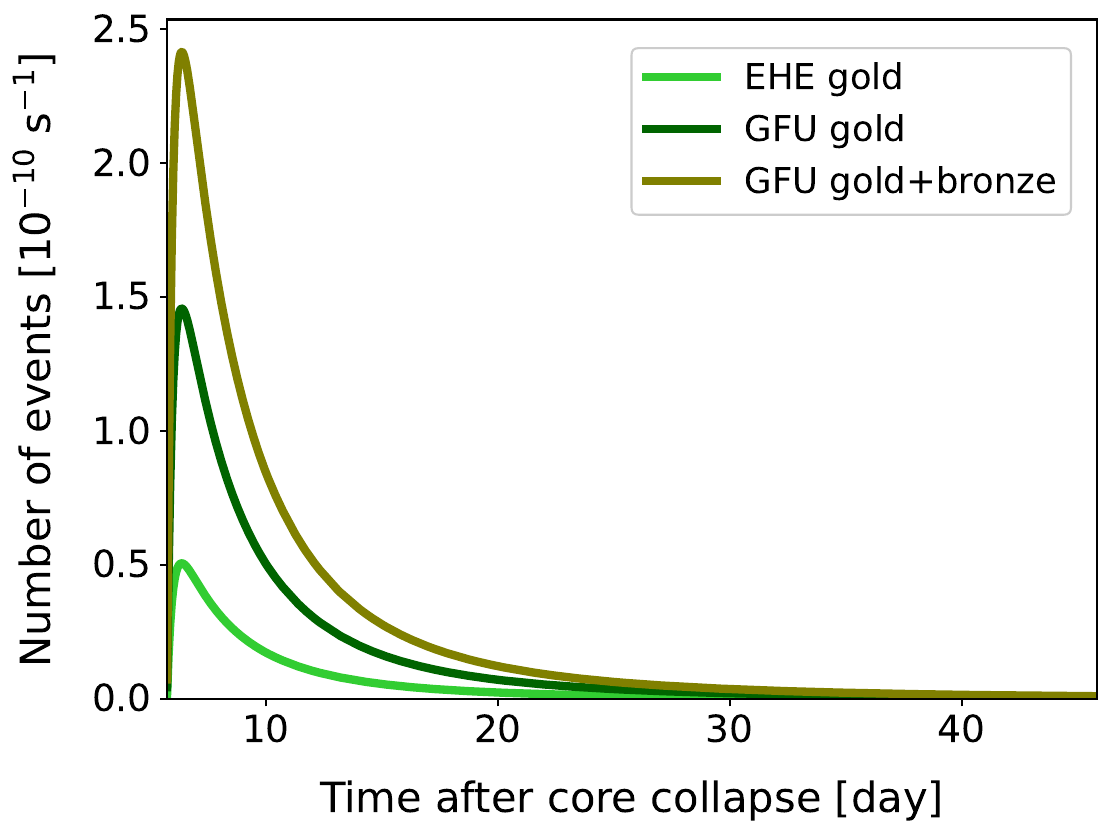}
    \caption{Top: the expected differential detected event rates at IceCube at different time slices based on the GFU gold+bronze effective area for the declination angle of 0 to 30~deg, with the same color notation as Figure~\ref{fig:elumi}. 
    Bottom: the expected number of detection events at IceCube along time for three different effective areas and the total detection number for each case shown in the legend. Here, only muon-type neutrinos are considered in both panels, as described in the main text.}
    \label{fig:nevent}
\end{figure}

Equation \eqref{eq:dNdt} defines a time-dependent expected event rate, $\lambda(t)\equiv dN_{\rm event}/dt$, which is the quantity directly comparable to detector expectations. 
In the high-statistics limit, where $\int \lambda(t)dt\gg1$, the observed time-differential signal would roughly trace this temporal profile. 
In the low-count regime relevant here, however,  $\int \lambda(t)dt\ll1$ and a detection, if it occurs, is expected to be rare. 
In this case, $\lambda(t)$ is best interpreted as a weighting function for when a rare detection is most likely. 
Conditioned on detecting a single event, the detection-time probability density is given by the normalized rate,
\begin{equation}
    p(t|1\,\mathrm{event})=\cfrac{\lambda(t)}{\int \lambda(t')dt'}~.
    \label{eq:nomal}
\end{equation}

Figure \ref{fig:nevent} summarizes the resulting IceCube expectations in energy and time. 
The top panel shows the expected differential detected event rates as a function of neutrino energy at representative epochs (computed using the GFU gold+bronze effective area for  0 to 30 deg declination), enabling a direct comparison with the reported energy of IC-231004A ($\sim$442 TeV). 
The bottom panel shows $\lambda(t)= dN_{\rm event}/dt$ for three effective-area selections (EHE gold, GFU gold, and GFU gold+bronze), 
and therefore provides, via the normalization in Eq. \eqref{eq:nomal}, the implied detection-time weighting (or probability density) in the low-count limit.

For a single SN~2023uqf-like event at $d = 723$~Mpc, our detector-folded calculation yields an expected number of IceCube track-like events of $2.0\times10^{-5}$ to $9.7\times10^{-5}$, depending on the alert selection (Table \ref{tbl:count}). 
These values are far below unity, so the result should be viewed as an event-specific consistency test rather than a statistical association claim. 
The detector-folded calculation quantifies, conditional on a rare detection, the relative timing and energy weighting of the event.

The broader implication should also be interpreted in the context of the Type Ibn population. 
As illustrated by the comparison with the Type Ibn sample in Figure~\ref{fig:em_LC}, 
SN 2023uqf lies on the luminous and rapidly evolving side of the class, but it is not an obvious photometric outlier. 
At the same time, the RHD model requires a dense CSM normalization, \(D'=50\), so SN 2023uqf should be regarded as an example of the dense-CSM subset of Type Ibn events rather than as a representative of the entire population.

For an event with similar optical luminosity, shock velocity, and CSM interaction properties, the expected number of IceCube track-like events scales approximately as 
\(\ N_\mathrm{event}\propto d^{-2}. \) 
Thus, while the expected number from any single event at \(d=723\) Mpc is small, the observed rate of Type Ibn supernovae \citep{2025A&A...703A..34P,2025A&A...698A.305M} and the strong \(d^{-2}\) scaling imply that nearby luminous and rapidly evolving Type Ibn events with comparable shock-dissipation histories are good candidates for targeted optical--neutrino searches.

This motivates continued searches in larger optical--neutrino samples. 
ZTF has been operating since 2018 and the IceCube archival data based on the event selections used in the currently-referred alert system exist since 2011~\citep{2023ApJS..269...25A}. 
In \citet{2025arXiv250808355S}, the optical-neutrino coincidence search was performed on archival data from the overlapped $\sim$6 years.
However, a quantitative assessment of their cumulative contribution to the total diffuse neutrino flux measured by IceCube remains a subject for future work, since it requires the distribution of CSM density, radial structure, shock velocity, and microphysical parameters across the Type Ibn population.

\begin{table}[]
    \caption{
    Expected number of IceCube track-like ($\nu_\mu + \bar{\nu}_\mu$) events from SN 2023uqf at $d = 723$ Mpc for different IceCube alert selections. 
    }
    
    \centering
    \begin{tabular}{l c} \hline 
        IceCube alert type & Expected detection number \\ \hline 
        EHE gold & $1.99 \times 10^{-5}$ \\ 
        GFU gold & $5.80 \times 10^{-5}$ \\ 
        GFU gold+bronze & $9.74 \times 10^{-5}$ \\ \hline
    \end{tabular}
    \label{tbl:count}
\end{table}

\begin{figure*}
    \centering
    \includegraphics[width=0.8\textwidth]{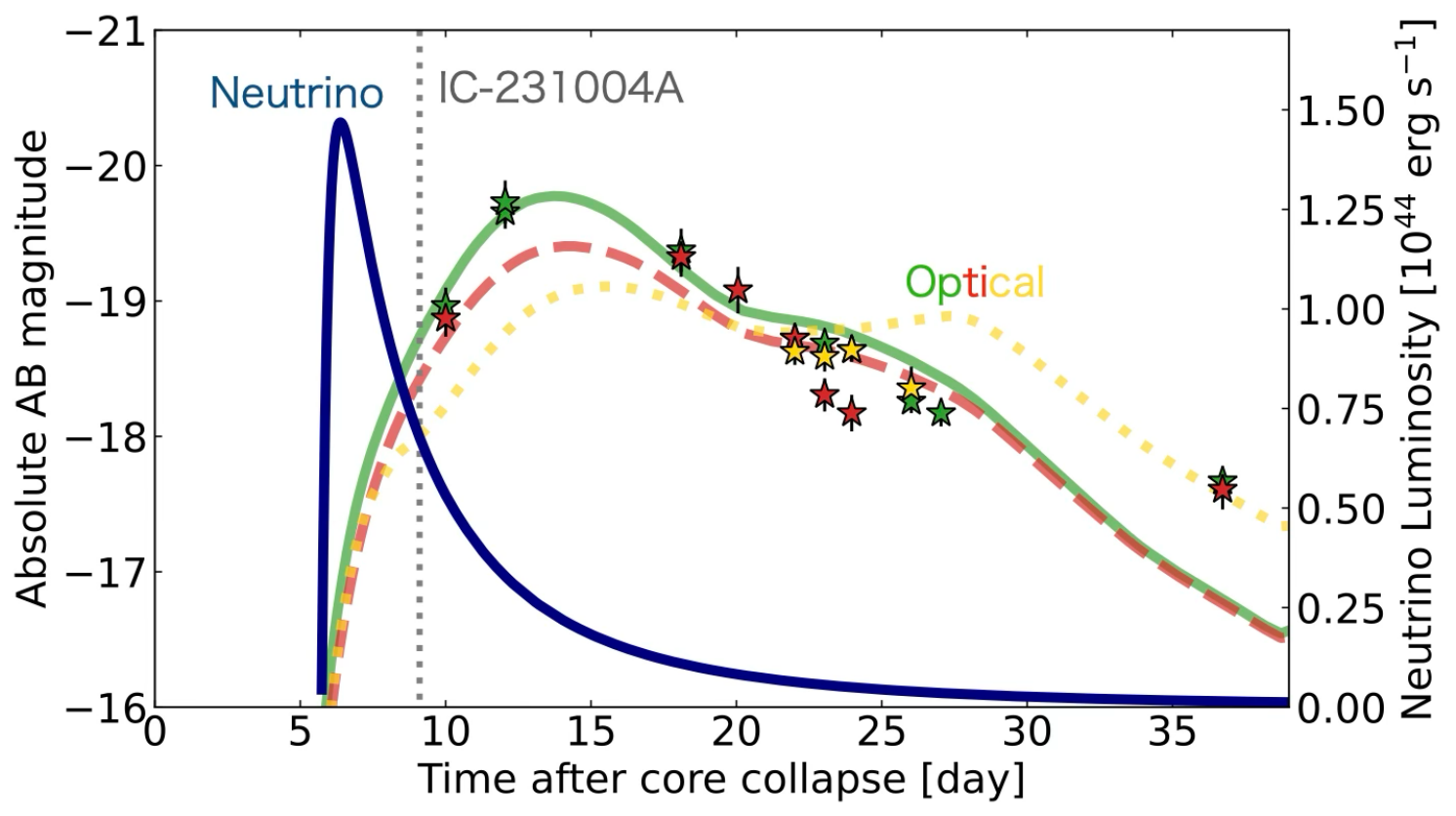}
    \caption{
    \textbf{Unified temporal picture of SN 2023uqf and IC-231004A.} 
    The colored curves and points show the optical light curves of SN 2023uqf (left axis; absolute AB magnitude), where the points are the ZTF measurements reported by \citet{2025arXiv250808355S} and the curves are our {\tt STELLA}-based optical model (Section \ref{sec:mm_em}). 
    The blue curve shows the modeled neutrino luminosity evolution from the shock–CSM interaction phase (right axis; summed over flavors; Section \ref{sec:mm_hinu}). The vertical dashed line marks the detection time of IC-231004A. 
    The explosion epoch inferred from the optical light-curve fit corresponds to $\Delta t\approx -10$ days on this axis (Section \ref{subsec:init_opt}), and the neutrino-luminosity curve is shifted accordingly.
    In the low-count regime, the time dependence of the expected IceCube event rate provides a weighting for when a rare detection is most likely; IC-231004A occurs within the high-weight time window implied by the model.
    }
    \label{fig:multi_LC}
\end{figure*}

\section{Discussion and Conclusion} \label{sec:discus}

We have investigated whether the reported optical-neutrino proximity between the Type Ibn supernova SN 2023uqf and the $\sim$442 TeV IceCube alert IC-231004A can be understood within a single, physically motivated shock–CSM interaction scenario. 
By combining (i) radiation-hydrodynamic light-curve constraints on the CSM density and shock evolution (Section \ref{sec:mm_em}), (ii) time-dependent hadronic neutrino production modeling (Section \ref{sec:mm_hinu}), and (iii) detector-folded event rate estimates in the low-count regime (Section \ref{sec:mm_nuobs}), we are able to assess whether the optically inferred shock--CSM environment can satisfy the energy and timing requirements implied by IC-231004A in a self-consistent framework.

A central outcome is summarized in Figure \ref{fig:multi_LC}, which places the optical emission, the modeled neutrino luminosity evolution, and the IC-231004A detection time on a common temporal axis. 
In the low-count regime relevant here, the detector-folded event rate should be interpreted as a weighting function for when a rare detection would be most likely, not as an association probability. 
The detection time of IC-231004A falls within the high-weight interval implied by this distribution, while the event energy scale is simultaneously compatible with the escape-limited modeled neutrino spectrum (Figure \ref{fig:nevent}, top). 
Taken together, these energetic and temporal consistencies show that
SN 2023uqf satisfies the basic optical-to-neutrino consistency requirements in our fiducial interaction model. 
This does not establish SN 2023uqf as the origin of IC-231004A, but it demonstrates that the optically inferred shock--CSM environment can, in principle, provide the timing and energy scale required for such a neutrino event.

If IC-231004A were associated with SN 2023uqf, its \(\sim442\) TeV energy would require parent hadrons with at least multi-PeV energies.  In our escape-limited fiducial model, 
SN 2023uqf highlights the early shock–CSM interaction phase as a plausible transient PeVatron window, in which collisionless acceleration can operate while the CSM density remains sufficiently large for efficient hadronic interactions. 
This interpretation is consistent with recent theoretical work arguing that interacting supernovae can accelerate nuclei to (super-)PeV energies during ejecta–CSM interaction \citep{2026arXiv260206410E}, while our study provides an event-specific, optical-anchored consistency test for SN 2023uqf and IC-231004A.

Several caveats may delimit the strength of our conclusions. 
The maximum energy of parent hadrons in our fiducial calculation is escape-limited rather than fully loss-limited, as discussed in Section \ref{subsec:cr}, and a more complete treatment of energy losses will be needed for precise spectral-cutoff predictions.
The neutrino yield depends on microphysical parameters (e.g., $\epsilon_p$, $\epsilon_B$) and on the CSM structure and ionization state, which affect both shock conditions and absorption at X-ray/radio wavelengths. 
We also note that the progenitor channel of Type Ibn supernovae may be diverse \citep[e.g.,][]{2025MNRAS.541.3748K,2025PASJ...77.1385M}.
In addition, our treatment of nuclear primaries and the use of simplified hadronic production kernels introduce modeling systematics at the factor-of-few level. 
Future progress will therefore benefit from (i) earlier and deeper X-ray/radio follow-up to constrain CSM ionization and shock emergence, (ii) refined explosion-epoch constraints to sharpen the timing test, and (iii) continued IceCube alerts as well as additional ones from other neutrino telescopes, such as KM3NeT~\citep{2024EPJC...84..885K} and Baikal~\citep{2021Symm...13..377S}, enabling comparisons of observed event times between optical and neutrinos. 
With these developments, interaction-powered supernovae may become a quantitatively testable component of the high-energy neutrino sky.

\begin{acknowledgments}
The work has been supported by Japan Society for the Promotion of Science (JSPS) 21K13964 and 26K17189 (RS).
Y.I. acknowledges financial support from Grant-in-Aid for the JSPS Fellows (25KJ1472).
Y.A. thanks support from the 2026 Inamori Research Grants (“Incubate” Course).
\end{acknowledgments}

\begin{contribution}
%%This section gives authors the space to recognize author contributions. The text inside this environment is NOT counted towards the total word quanta. At a minimum, manuscripts are expected to include this text:
All authors contributed equally to this work.
%% Authors can use the Contributor Role Taxonomy (CRediT) at
%% https://credit.niso.org
%% for ideas on how write a good statement tailored to their needs.
\end{contribution}

%% To help institutions obtain information on the effectiveness of their telescopes the AAS Journals has created a group of keywords for telescope facilities.
%\facilities{}

\software{{\tt STELLA} \citep{1998ApJ...496..454B,2000ApJ...532.1132B,2006A&A...453..229B}}

%\appendix

%\restartappendixnumbering
%\section{Appendix information}\label{sec:app}

\bibliography{ref}{}
\bibliographystyle{aasjournal}

\end{document}